\documentclass{sig-alternate}

\usepackage{url}
\graphicspath{{figs/}}
\DeclareGraphicsExtensions{.pdf,.jpeg,.png}
\usepackage{array}
\usepackage{multirow}
\usepackage{tabulary}
\usepackage{subcaption}
\usepackage[caption=false]{subfig}

\newcommand{\gap}{{\boldsymbol \star}}
\newlength{\aligncharw}
\newcommand{\agap}{\makebox[\aligncharw][s]{\scriptsize{$\gap$}}}

\newcommand{\ie}{i.e.\,}
\newcommand{\cf}{cf.\,}

\newcommand{\ScaleF}{\mbox{$\sigma$}}
\newcommand{\Real}{\mbox{$\mathbb{R}$}}
\newcommand{\MsgChar}{\mbox{$\mathcal{C}$}}
\newcommand{\MessageSet}{\mbox{$\mathcal{M}$}}

\newcommand{\Request}{\mbox{\it Rq}}

\newcommand{\Response}{\mbox{\it Rp}}

\newcommand{\Interaction}{\mbox{$I$}}
\newcommand{\InteractionTrace}{\mbox{$I_1I_2I_3\ldots I_n$}}
\newcommand{\InteractionSet}{\mbox{$\mathcal{I}$}}
\newcommand{\InteractionLib}{\mbox{$I^*({\InteractionSet})$}}
\newcommand{\Pair}[2]{\mbox{$(#1, #2) $}}

\newcommand{\DistFunc}{\mbox{$dist$}}
\newcommand{\Score}{\mbox{$score$}}
\newcommand{\TransFunc}{\mbox{$trans$}}

\begin{document}

% Copyright
%\setcopyright{acmcopyright}
%\setcopyright{acmlicensed}
%\setcopyright{rightsretained}
%\setcopyright{usgov}
%\setcopyright{usgovmixed}
%\setcopyright{cagov}
%\setcopyright{cagovmixed}

%\CopyrightYear{2016} 
%\setcopyright{acmlicensed}
%\conferenceinfo{CSED'16,}{May 14-15 2016, Austin, TX, USA}
%\isbn{978-1-4503-4157-8/16/05}\acmPrice{\$15.00}
%\doi{http://dx.doi.org/10.1145/2896941.2896950}

% DOI
%\doi{10.475/123_4}

% ISBN
%\isbn{123-4567-24-567/08/06}

%Conference
%\conferenceinfo{PLDI '13}{June 16--19, 2013, Seattle, WA, USA}

%\acmPrice{\$15.00}

%
% --- Author Metadata here ---
%\conferenceinfo{WOODSTOCK}{'97 El Paso, Texas USA}
%\CopyrightYear{2007} % Allows default copyright year (20XX) to be over-ridden - IF NEED BE.
%\crdata{0-12345-67-8/90/01}  % Allows default copyright data (0-89791-88-6/97/05) to be over-ridden - IF NEED BE.
% --- End of Author Metadata ---

\title{Enhanced Playback of Automated Service Emulation Models Using Entropy Analysis}

\numberofauthors{6} %  in this sample file, there are a *total*
\author{
% 1st. author
\alignauthor
Steve Versteeg\\
       \affaddr{CA Research}\\
       \affaddr{CA Technologies}\\
       \affaddr{Melbourne, Australia}\\
       \email{steve.versteeg@ca.com}
% 2nd. author
\alignauthor
Miao Du\\
       \affaddr{Swinburne University of Technology}\\
       \affaddr{Hawthorn, Victoria, Australia}\\
       \email{miaodu@swin.edu.au}
% 3rd. author
\alignauthor John Bird\\
       \affaddr{CA Technologies}\\
       \affaddr{Melbourne, Australia}\\
       \email{john.bird@ca.com}
\and
\alignauthor Jean-Guy Schneider\\
       \affaddr{Swinburne University of Technology}\\
       \affaddr{Hawthorn, Victoria, Australia}\\
       \email{jschneider@swin.edu.au}
\alignauthor John Grundy\\
       \affaddr{Deakin University}\\
       \affaddr{School of Info.\! Technology}\\
       \affaddr{Burwood, Victoria, Australia}\\
       \email{j.grundy@deakin.edu.au}
\alignauthor Jun Han\\
       \affaddr{Swinburne University of Technology}\\
       \affaddr{Hawthorn, Victoria, Australia}\\
       \email{jhan@swin.edu.au}
%\and  % use '\and' if you need 'another row' of author names
% 4th. author
%\alignauthor Lawrence P. Leipuner\\
%       \affaddr{Brookhaven Laboratories}\\
%       \affaddr{Brookhaven National Lab}\\
%       \affaddr{P.O. Box 5000}\\
%       \email{lleipuner@researchlabs.org}
%% 5th. author
%\alignauthor Sean Fogarty\\
%       \affaddr{NASA Ames Research Center}\\
%       \affaddr{Moffett Field}\\
%       \affaddr{California 94035}\\
%       \email{fogartys@amesres.org}
%% 6th. author
%\alignauthor Charles Palmer\\
%       \affaddr{Palmer Research Laboratories}\\
%       \affaddr{8600 Datapoint Drive}\\
%       \affaddr{San Antonio, Texas 78229}\\
%       \email{cpalmer@prl.com}
}
% There's nothing stopping you putting the seventh, eighth, etc.
% author on the opening page (as the 'third row') but we ask,
% for aesthetic reasons that you place these 'additional authors'
% in the \additional authors block, viz.
%\additionalauthors{Additional authors: John Smith (The Th{\o}rv{\"a}ld Group,
%% Just remember to make sure that the TOTAL number of authors
% is the number that will appear on the first page PLUS the
% number that will appear in the \additionalauthors section.

\maketitle
\begin{abstract}
  Service virtualisation is a supporting tool for DevOps to generate
  interactive service models of dependency systems on which a
  system-under-test relies. These service models allow applications under
  development to be continuously tested against production-like
  conditions. Generating these virtual service models requires expert
  knowledge of the service protocol, which may not always be
  available. However, service models may be generated automatically from
  network traces. Previous work has used the Needleman-Wunsch algorithm to
  select a response from the service model to play back for a live request. We
  propose an extension of the Needleman-Wunsch algorithm, which uses entropy
  analysis to automatically detect the critical matching fields for selecting
  a response. Empirical tests against four enterprise protocols demonstrate
  that entropy weighted matching can improve response accuracy.
\end{abstract}

%\begin{CCSXML}
%<ccs2012>
%  <concept_id>10010520.10010553.10010562</concept_id>
%  <concept_desc>Computer systems organization~Embedded systems</concept_desc>
%  <concept_significance>500</concept_significance>
%  <concept_id>10010520.10010575.10010755</concept_id>
%  <concept_desc>Computer systems organization~Redundancy</concept_desc>
%  <concept_significance>300</concept_significance>
%  <concept_id>10010520.10010553.10010554</concept_id>
%  <concept_desc>Computer systems organization~Robotics</concept_desc>
%  <concept_significance>100</concept_significance>
%  <concept_id>10003033.10003083.10003095</concept_id>
%  <concept_desc>Networks~Network reliability</concept_desc>
%  <concept_significance>100</concept_significance>
%</ccs2012>  
%\end{CCSXML}

%\ccsdesc[500]{Computer systems organization~Embedded systems}


%
%  Use this command to print the description
%
%\printccsdesc

% We no longer use \terms command

%\keywords{ACM proceedings; \LaTeX; text tagging}

\section{Introduction}
\label{sec:intro}

Continuous delivery is an emerging practice in Software Engineering which aims
to compress the release cycle such that a completed developer change can be
quickly released (within a timeframe of hours) to the end-user. It aims to
satisfy the business demand for agility and can be applied to both Cloud
computing and on-premise software.

Continuous delivery relies on automating all stages of the release cycle,
including software builds, unit testing, integration testing, performance
testing and end-user environment testing. For continuous delivery, fully
automated tests need to be performed in ``production like conditions''
\cite{humble2010continuous}. For on-premise enterprise software, this is
particularly challenging. Each enterprise environment is unique. Furthermore, a
software system-under-test (SUT) will be integrated with other systems, such
that its behaviour depends on the interactions it makes with the other
systems. As a consequence, upgrading, replacing or installing a new service in
an enterprise environment is high risk.
%Performing production like testing for
%an enterprise environment, therefore, requires an accurate replication of the
%dependency services on which the SUT relies.
To be confident the SUT can successfully operate in production, one needs
an accurate replication of the real dependency services' behaviours (including any bugs),
at their current versions and configurations.
The traditional approach includes having a test environment which
is a close replication of the production environment. This is not
only expensive to build and maintain, but is difficult
to automate making, this approach incompatible with continuous delivery.

Service virtualisation \cite{Michelsen:12} (also known as service emulation)
is a technology to build accurate interactive models of the dependency
services on which a system-under-test relies. This is achieved by
recording on-the-network interactions between a system-under-test and other
services in the production environment, and using this as the basis
for building a virtualised service model.
Since the models are built directly from the real dependency
services, their interactive behaviour may be quite accurate.
Furthermore, service models are conducive to automation, as they
can be easily distributed and incorporated into automated tests.
IT operations staff, quality assurance teams and application developers,
may all make use of virtual service models to test software in
production like conditions.
Service virtualisation is, therefore, a
critical tool for accelerating the software release cycle and
supporting the ultimate goal of continuous delivery.

Service virtualisation requires a Data Protocol Handler (DPH)
for every dependency service on which the system-under-test relies.
The DPH processes the syntax and semantics of recorded network messages --
headers are identified, operations and arguments are
extracted, and business rules are identified.
Service virtualisation tools
provide a set of predefined DPHs covering the most commonly used protocols.
However, some of the dependency services may use a less common, and therefore
unsupported, protocol. Examples of this include proprietary systems,
legacy systems, specialised domain applications, custom built or in-house applications,
and mainframe systems. Yet for the system-under-test to be able to operate in a virtualised
environment, it is essential that all of the dependency systems be virtualised.
If even one of the dependency systems is missing, production like test can, in
general, not be performed.

For any unsupported protocol, a custom DPH needs to be written.
This may require extensive hours of programming as well as detailed knowledge
of the target protocol, provided by thorough documentation and application specific
knowledge given by the application architect.
In practice such detailed knowledge is
often unavailable or incomplete. For example, documentation may be missing,
not updated, or the original application architect may have left the
organisation.
Without the required knowledge, writing the custom DPH becomes
impossible, causing the entire service virtualisation project to fail.

To address this gap, an alternative method for service virtualisation was
proposed~\cite{Du:2013}, the method, dubbed \emph{opaque service
  virtualisation}, uses the Needleman-Wunsch algorithm \cite{needleman:1970}
to perform byte-level matching, when selecting a response from the opaque
service model to replay. This method assumes no knowledge of the target
protocol, enabling a service to be modelled automatically, even in the absence
of the expert knowledge that would otherwise be required.

The previous opaque service virtualisation technique~\cite{Du:2013} has limited accuracy
when a response of the wrong operation type is replayed. Since there is no knowledge of
the protocol or message structure, there is no way to prioritise matching the operation type over other parts
of the message (\ie the payload).
%
%%% Merged paragraphs (JGS) %%%
%
We propose an extension to opaque service virtualisation, whereby entropy analysis is used
is used to prioritise matching responses for the correct operation type. The technique
extends the Needleman-Wunsch algorithm to perform an entropy weighted match.

\section{Related Work}
\label{sec:relatedwork}

The most common approach to recreating a production like environment for a system-under-test is to use virtual machines \cite{li:10}. Implementations of the services are deployed on virtual machines and communicated with by the system under test. Major challenges with this approach include configuration complexity \cite{vm4testing} %, limited scalability
 and the need to maintain instances of each and every service type in multiple configurations \cite{grundy:05}. Recently, cloud-based testing environments~\cite{testenvcloud} as well as containerisation~\cite{docker} have emerged to mitigate some of these issues.

Emulated testing environments for enterprise systems, relying on service
models, is another approach. When sent messages by the system
under test, the emulation responds with approximations of ``real'' service response
messages \cite{servicemodel}. Kaluta \cite{hine:thesis} is proposed to
provision emulated testing environments. Challenges with these approaches
include developing the models, lack of precision in the models, especially
for complex protocols, and ensuring robustness of the models under diverse
load conditions \cite{sun2012usefulness}.

%  To assist developing reusable
%service models, approaches either reverse engineer message structures
%\cite{cui:07a}, or discover processes \cite{processmining}. % for building
%%behavioral models \cite{behavioralmodel}. 
%While these allow engineers to
%develop more precise models, none of them can automate the creation
%of executable interactive models of the communication between
%a system under test and the dependency services.

Recording and replaying message traces is an alternative approach. This involves recording request messages sent by the system under test to real services and response messages from these services, and then using these message traces to `mimic' the real service response messages in the emulation environment \cite{cui:06}. Some approaches combine record-and-replay with reverse-engineered service models. CA Service Virtualization~\cite{Michelsen:12} is a commercial software tool, which can emulate the behaviour of services. The tool uses built-in knowledge of some protocol message structures to model services and mimic interactions automatically.

Roadmaps of research challenges relating to continuous delivery broadly have also been proposed \cite{chen2015continuous,fitzgerald2014continuous}.  
Entropy analysis has been used in other domains, for example in anti-malware research, entropy profiles have been used to classify packers \cite{ebringer2008fast}.

%Active Trace Clustering for Improved Process Discovery - De Weerdt, vanden Broucke, Vanthienen, Baesens
%
%Process diagnostics using trace alignment: Opportunities, issues and challenges - Jagadeesh Chandra Bose, van der Aalst
%
%LISA
%
%Kaluta
%
%State of the art in sequence alignment

\section{Response Playback Framework}

\begin{table}[t!]
\begin{center}
\resizebox{0.5\textwidth}{!}{
\begin{tabular}{|c||l|l|}
\hline
\# & Request & Response \\ \hline\hline
1 & \{id:001,op:S,sn:Du\} & \{id:001,op:SearchRsp,result:Ok,\\
  &                       & ~~gn:Miao,sn:Du,mobile:5362634\} \\ \hline
2 & \{id:013,op:S,sn:Versteeg\} & \{id:013,op:SearchRsp,result:Ok, \\
  &                             & ~~gn:Steve,sn:Versteeg,mobile:9374723\} \\ \hline
3 & \{id:024,op:A,sn:Schneider\} & \{id:024,op:AddRsp,result:Ok\} \\ \hline
4 & \{id:275,op:S,sn:Han\} & \{id:275,op:SearchRsp,result:Ok, \\
  &                        &   ~~gn:Jun,sn:Han,mobile:33333333\} \\ \hline
5 & \{id:490,op:S,sn:Grundy\} & \{id:490,op:SearchRsp,result:Ok,\\
  &                           & ~~gn:John,sn:Grundy,mobile:44444444\} \\ \hline
6 & \{id:773,op:S,sn:Hine\} & \{id:273,op:SearchRsp,result:Ok, \\
  &                         & ~~sn:Hine,mobile:123456\} \\ \hline
7 & \{id:887,op:A,sn:Will\} & \{id:887,op:AddRsp,result:Ok\} \\ \hline
8 & \{id:906,op:A,sn:Hine\} & \{id:906,op:AddRsp,result:Ok\} \\
\hline
\end{tabular}
}
\end{center}
\caption{Directory Service Interaction Library Example}
\label{tab:tl}
\end{table}

Opaque service virtualisation consists of two parts:
\begin{enumerate}
\item Record an \emph{interaction library} -- 
a sample set of messages exchanges between a system-under-test
and a dependency service (the \emph{target service}). These messages may
be collected using a network analyser tool, such as WireShark, or logged
through a proxy.
\item Deploy an emulation of the target service. The emulator receives
live requests on the network from the system-under-test. The emulator
searches the interaction library for the nearest matching request,
and plays back the corresponding response.
\end{enumerate}

We will now give a playback example and then describe in detail the
previous approach to the response selection step.

\subsection{Non-Weighted Playback Example}
\label{ss:noweightexample}

Table~\ref{tab:tl} shows a small example interaction library. It is from a
fictional directory service protocol that has some similarities to the widely
used LDAP protocol~\cite{ldap}, but is simplified to make our running example
easier to follow. Our example protocol uses a JSON encoding. The transaction
library contains two kinds of operations: {\sf {\small add}} and {\sf {\small
    search}}. Add requests contain the field {\tt op:A}, whereas search requests
have {\tt op:S}. The add and search response operations are specified with the
fields {\tt op:AddRsp} and {\tt op:SearchRsp}, respectively.

Suppose we receive a live search request, for example, 
{\tt \{id:552,op:S,sn:Hossain\}}. Using the Needleman-Wunsch response
selection method~\cite{Du:2013}, then request 4 ({\tt \{id:275,op:S, sn:Han\}}, \DistFunc=0.125)
is the nearest request in the interaction library and the emulator will
replay a search response. For this case the behaviour is correct.

Now suppose we receive another live search request:\\ {\tt \{id:024,op:S,sn:Schneider\}}.
When selecting a response, request 3 ({\tt \{id:024,op:A,sn:Schneider\}},
\DistFunc=0.019) is a nearer match than request 4 ({\tt \{id:275,op:S,sn:Han\}},
\DistFunc= 0.15), even though the latter is a search request and the former is
an add request. Due to there being no prioritisation for matching the
operation type characters ({\tt op:A}) over the payload characters, a response
of the wrong operation type may be played back.

\subsection{Definitions}
\label{ss:defn}

We define a number of constructs needed to
express our framework more formally. We start with the notion of the most
basic building block, the set of {\em message characters}, denoted by
{\MsgChar}. We require equality and inequality to be defined for the elements of
{\MsgChar}. For the purpose of our study, {\MsgChar} could comprise
of a set of valid bytes that can be transmitted over a network or characters from a
character set (such as ASCII or Unicode) or the set of
printable Characters as a dedicated subset. Furthermore, we define
{\MessageSet} to be the set of all (possibly empty) {\em messages} that can be
defined using the message characters. A message $m \in \MessageSet$ is a
non-empty, finite sequence of {message characters} $c_1c_2c_3\ldots c_n$ with
$c_i \in \MsgChar, 1 \leq i \leq n$.
%
%%% May not be neeeded...
%We consider two messages $m_1 = c_{1,1}c_{1,2}\ldots c_{1,l}$
%and $m_2= c_{2,1}c_{1,2}\ldots c_{2,n} $ to be equal if $l=n$ and $c_{1,i} =
%c_{2,i}, \forall\; 1 \leq i \leq n$.

Without loss of generality, we assume that each request is always followed by
a single response. If a request does not generate a response, we insert a
dedicated ``no-response'' message into the recorded interaction traces. If, on
the other hand, a request leads to multiple responses, these are concatenated
into a single response.

A single interaction {\Interaction} consists of a request, denoted by
{\Request}, as well as the corresponding response, denoted by
{\Response}. Both {\Request} and {\Response} are elements of {\MessageSet} and
we write {\Pair{\Request}{\Response}} to denote the corresponding
request/response pair.

An {\em interaction trace} is defined as a finite,
non-empty sequence of interactions, that is, {\InteractionTrace}. Finally,
we define the set of interactions {\InteractionSet} as a non-empty set of
interaction traces.

%% For the purpose of our proposed technique as described above, we assume
%% that for a given protocol under investigation, we are able to record a
%% number of sufficiently large interactions between two (or more) software
%% endpoints.  Tools like Wireshark \cite{lamping:12a} have the necessary
%% functionality to filter network traffic and record messages of interest in
%% a suitable format for further processing. We also assume that these
%% recordings are ``valid'', that is, that the sequence of interactions in our
%% recorded interaction trace database are correct with regards to the
%% temporal properties of the underlying protocol and that each request and
%% response message is well-formed.

% 3. Request/Resonse pairs
% 4. Interaction trace -> ordered sequence of Request/Response pairs

%We define a software service in a broad sense to mean any program with a callable API, invokable by
%some client application. The API will typically be called over the network, but could
%be invoked via memory. Some example services include a RESTful web service, a SOAP web service, a directory server or a mainframe program.

\subsection{Non-Weighted Response Selection}

Our previous framework consisted two main processing steps:
(i) given an incoming request from the system-under-test to a virtual service model,
we search for a suitably ``similar'' request in the
previously recorded interaction traces. (ii) Our system then synthesizes a
response for the incoming request based on the similarities in the request
itself and the ``similar'' request identified in the interaction traces, as
well as the recorded response of the ``similar'' request.

Using the definitions introduced above, our framework can thus be formalized
as below. To facilitate the presentation, we denote ${\Request}_{in}$ as the
incoming request and the interaction library $I^*({\InteractionSet})$ as the set of all interactions
in {\InteractionSet}.

%%% I will try a few formatting tricks here...
{\hspace{0.4cm}}
${\Response}_{out} = \mathbf{\TransFunc} \ ({\Request}_{in}, {\Request}_{sim},
      {\Response}_{sim})$

with {\vspace{-0.23cm}}
\begin{itemize}
\item ${\Pair{{\Request}_{sim}}{{\Response}_{sim}}} \in I^*({\InteractionSet})$;
and \\[-0.60cm]
\item $\forall \; {\Pair{{\Request}_{i}}{{\Response}_{i}}} :
{\DistFunc}({\Request}_{in}, {\Request}_{sim}) \leq {\DistFunc}({\Request}_{in}, {\Request}_{i})$
\end{itemize}
where {\DistFunc} and {\TransFunc} denote user-defined {\em distance} and {\em
  translation} functions, respectively, allowing the framework to be tailored
for the specific needs of given context.

%% The distance function {\DistFunc} is used to compute the distance between two
%% requests. We solely require the distance of a message $m$ with itself
%% to be zero, that is ${\DistFunc}(m, m) = 0$. Particularly, {\DistFunc} does
%% not need to be symmetric, and the distance function based on common
%% subsequence matching used in our experiments is asymmetric. Depending on what
%% kind of distance function is used, a different pre-recorded request will be
%% chosen to be the most ``similar'' to the incoming request.

The distance function {\DistFunc} is used to compute the distance between two
requests. We require (i) the distance of a message $m$ with itself to be zero,
that is ${\DistFunc}(m, m) = 0$, and (ii) the distance between two
non-identical messages $m_1$ and $m_2$ to be greater than zero. Depending on
what kind of distance function is used, a different pre-recorded request will
be chosen to be the most ``similar'' to the incoming request.
We used the Needleman-Wunsch {\em edit distance} \cite{needleman:1970} as the
basis for {\DistFunc}.

The {\em translation} function's responsibility is to synthesize a response
for the incoming request. As a simplification of our work, we made the
decision to {\em ignore} temporal properties in our framework, that is, the
synthesized response solely depends on the incoming request and the recorded
interaction traces, but not on any previously received or transmitted requests
or responses, respectively. Adding a temporal dimension to the framework is
 part of our future work.

\subsection{Needleman-Wunsch}
\label{ss:nw}

The Needleman-Wunsch algorithm \cite{needleman:1970} is a dynamic programming
algorithm for computing the edit distance between two sequences.
Needleman-Wunsch finds the globally optimal alignment for two sequences of
symbols in $O(n_1\! \cdot\! n_2)$ time, where $n_1$ and $n_2$ are the lengths of the
sequences.

\begin{figure}[ht]
\centering
\begin{minipage}[b]{0.25\textwidth}
\includegraphics[width=.9\textwidth]{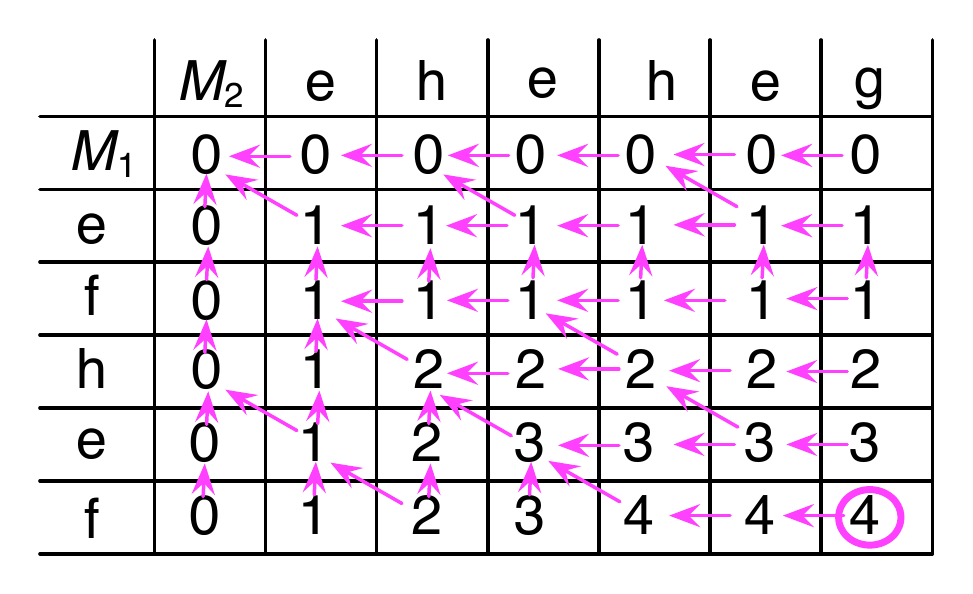}\\
\vspace{-.5cm}
\subcaption{$F$-matrix fill}
\label{fig:nwfill}
\end{minipage}%
\begin{minipage}[b]{0.25\textwidth}
\includegraphics[width=.9\textwidth]{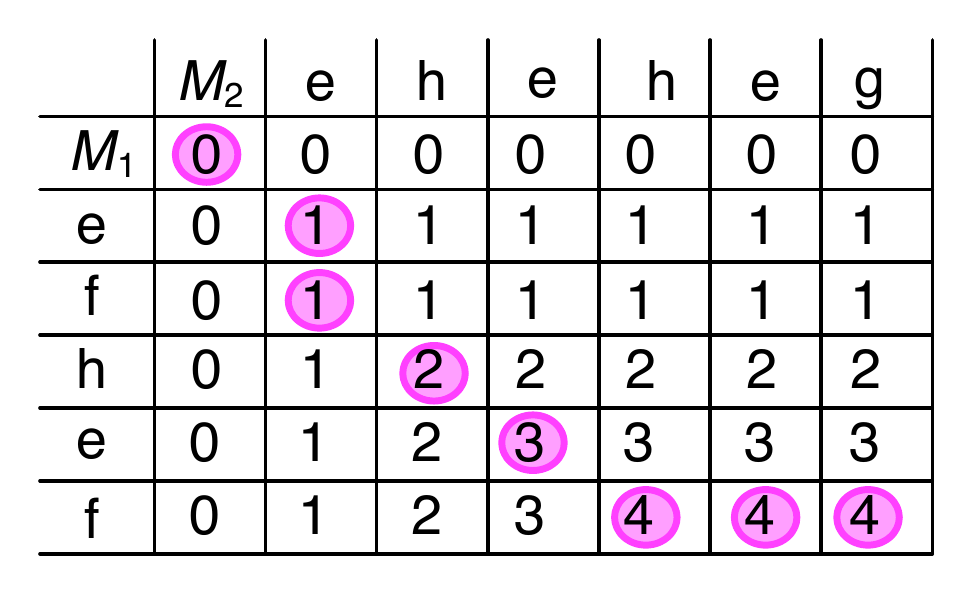}\\
\vspace{-.5cm}
\subcaption{$F$-matrix trace back}
\label{fig:nwtraceback}
\end{minipage}%
\caption{Needleman-Wunsch example of aligning $M_1 = \text{`efheh'}$ and $M_2 = \text{`eheheg'}$, using $d_{\rm identical}=1, d_{\rm differing} = -1, d_{\rm gap} = 0$.}
\label{fig:nwexample}
\end{figure}

Needleman-Wunsch progressively constructs a matrix $F$, using a scoring function $S(a, b)$ 
to score pairs of aligned symbols:
\begin{equation}
S(a,b) = 
\begin{cases}
d_{\rm identical}  &~\text{if}~a = b \\
d_{\rm differing} &~\text{otherwise}\\
\end{cases}
\end{equation}
where $d_{\rm identical}, d_{\rm differing}$ are constants. The constant $d_{\rm gap}$ is the gap penalty constant.

Needleman-Wunsch has three stages. To align two sequences $M_1$ (of length $n_1$) and $M_2$ (length $n_2$): 
\begin{enumerate}
\item An $(n_1+1) \times (n_2+1)$ matrix $F$ is constructed, with rows and columns labelled from 0..$n_1$ and 0..$n_2$, respectively. Row 0 and column 0 are initialised to 0.
\item $F$ is calculated progressively. At each cell: 
\begin{equation}
F_{i,j} = {\it max}
	\left\{ \begin{array}{l}
		F_{i-1,j-1} + S(M_1(i), M_2(j)) \\
		F_{i,j-1} + d_{\rm gap} \\
		F_{i-1,j} + d_{\rm gap} \\
	\end{array}
	\right.
	\label{eq:F}
\end{equation}
(See Figure~\ref{fig:nwfill} for example.)
\item $F_{n_1,n_2}$ gives $\Score(M_1, M_2)$, the optimal alignment score of $M_1$ and $M_2$.
The trace back step (see Figure~\ref{fig:nwtraceback}) captures the alignment.
\end{enumerate}

For example (in Figure~\ref{fig:nwexample}), the strings $M_1 = \text{`efheh'}$ and $M_2 = \text{`eheheg'}$ have an alignment score $\Score = 4$ and are aligned as:

\smallskip
\begin{tabular}{l}
\texttt{efheh{\agap\agap}} \\
\texttt{e{\agap}heheg} \\
\end{tabular}
\smallskip

To normalise for sequence length, we define the distance of $M_2$ relative to $M_1$ as:
\begin{equation}
\DistFunc(M_1, M_2) = \frac{\Score(M_1, M_2)-\Score_{\rm min}}{\Score_{\rm max} - \Score_{\rm min}}
\end{equation}
where $\Score_{\mathrm{max}}$ denotes the maximum possible alignment score
% for the given {\prototype}
and $\Score_{\mathrm{min}}$ the minimum possible alignment score for $M_1$
as defined by equations \ref{eq:maxscore} and \ref{eq:minscore},
respectively.

\begin{equation}
\label{eq:maxscore}
{\Score_{\mathrm{max}}(M_1)} = \sum_{i=1}^{|M_1|} S(M_1(i),M_1(i))
\end{equation}

\begin{equation}
\label{eq:minscore}
{\Score_{\mathrm{min}}(M_1)} = \sum_{i=1}^{|M_1|} S(M_1(i),\varnothing)
\end{equation}
where $\varnothing \not \in \MsgChar$ is a special symbol not in the character set.

% ------------------------------------------------------------------------ %

\section{Entropy Weighted Approach}

\begin{figure}
\centering
\includegraphics[width=0.45\textwidth]{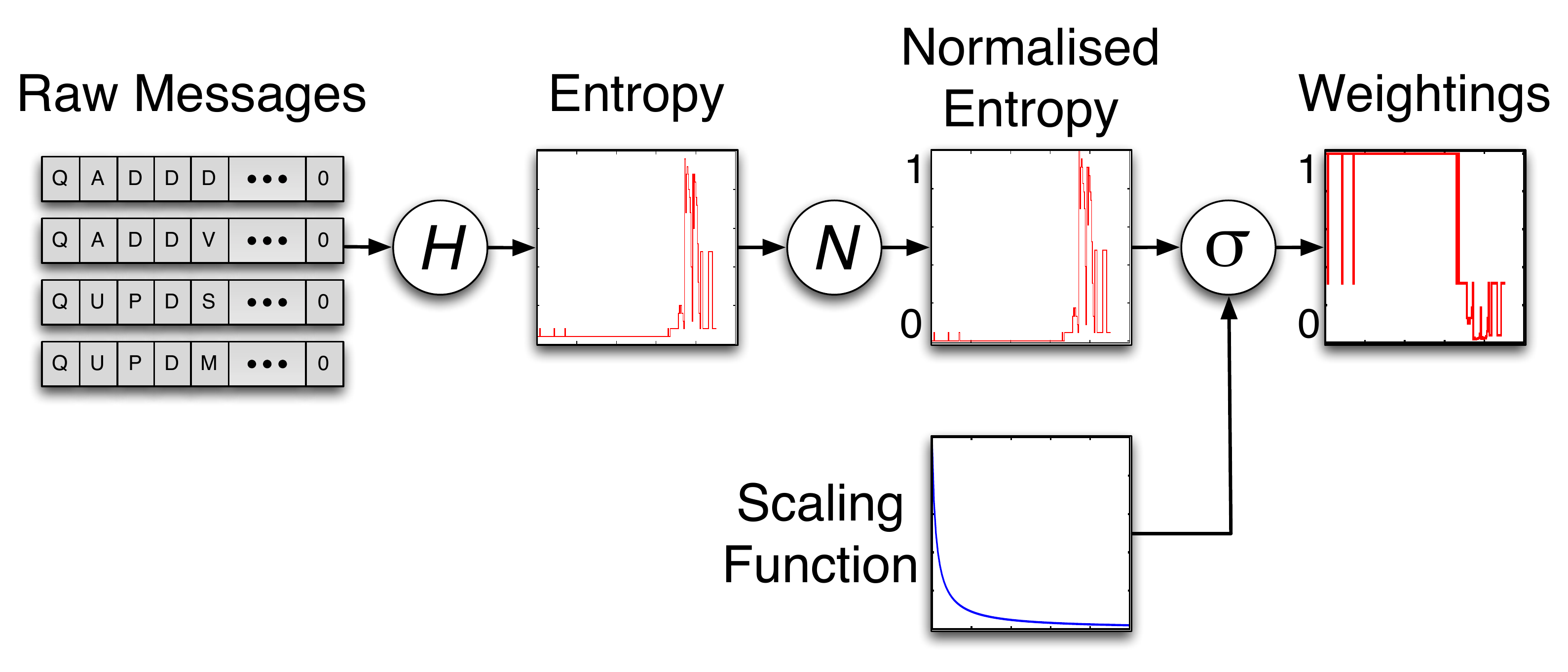}
\caption{Transformation steps in deriving a set of entropy weights}
\label{fig:entropycalc}
\end{figure}

We now propose a method of predicting which parts of a message contain the
operation type and other structural information, using entropy analysis. Our
method assumes no prior knowledge of the protocol message structure. We make
use of the observation that the parts of the message which contain operation
type and structural information are more stable (have less possibilities) than
the parts of the message containing payload information. We perform an entropy
analysis on the interaction library to derive a weightings vector, which is
applied to prioritise matching for the Needleman-Wunsch distance calculation.
The character positions with a low entropy (stable) are given a high
weighting, and character positions with high entropy (unstable) are given a
low weighting. Figure~\ref{fig:entropycalc} depicts an overview of our
approach.

Let $R = (r_{i,j})$ be a matrix of requests of normalised length:
\begin{equation}
r_{i,j} =
\begin{cases}
\Request_{i,j} & \text{if}\ j < |\Request_i| \\
\lambda & \text{otherwise}
\end{cases}
\end{equation}
where $\Request_{i,j}$ is the $j$th character of the $i$th request in the
interaction library \InteractionLib, $1 \leq i \leq |\InteractionLib|, 1 \leq
j \leq L$. $L = {\it max} \{ |\Request_i| \}$ is the length of the longest
request, and $\lambda \not\in \MsgChar$ is a special character to denote a
missing value.

Let $q_j(c)$ be the relative frequency of the character $c \in \MsgChar$ for
the $j$th column of $R$. Let the set $Q_j = \{q_j(c) : c \in \MsgChar\}$ be
the set of relative frequencies for all characters at the $j$th column
position of $R$.

To calculate a weightings vector, we require three functions:
\begin{itemize}
\item
Let $H: {\Real}^n \rightarrow {\Real}$ be a method for calculating the entropy
of the set of real numbers.
\item Let $N: {\Real} \rightarrow [0,1]$ be a normalisation function.
\item Let $\sigma: [0,1] \rightarrow [0,1]$ be a scaling function.
\end{itemize}

We define the weightings vector ${\bf w} = (w_1, w_2, ... w_{L})$, 
which gives a weighting for each column in the matrix $R$, with:
\begin{equation}
w_j = \sigma(N(H(Q_j)))
\end{equation}

\subsection{Entropy Measures}
\label{ss:entropymeasure}

We calculated the entropy of each column position of the matrix $R$,
considering three alternative entropy methods:

\begin{itemize}
\item
Shannon Index \cite{shannon:1948} is based on the weighted geometric mean of the relative frequencies of the characters, as given by Equation~\ref{eq:shannon}.
\begin{equation}
\label{eq:shannon}
H_{\rm Shannon}(Q) = -\sum^{|Q|}_{q_k \in Q} q_k \log q_k
\end{equation}

\item
Richness~\cite{jost2010relation} is a simple count of how many different characters occur at a give column, \ie 
\begin{equation}
\label{eq:richness}
H_{\rm Richness}(Q) = |Q'|
, \text{~where~} Q' = \{q_k : q_k \in Q \land q_k > 0\}
\end{equation}

\item
The Simpson Index~\cite{simpson1949measurement} is another measure of the concentration of types. It is defined as:
\begin{equation}
\label{eq:simpson}
H_{\rm Simpson}(Q) = \sum^{|Q|}_{q_k \in Q} q_k^2
\end{equation}

\end{itemize}

\subsection{Range Normalisation}

The entropy measures described in Section~\ref{ss:entropymeasure} have differing ranges.
We introduce a normalisation step, such that the entropy measured in each column of R is in the range [0,1]. Our normalisation function is:
\begin{equation}
N(x) = \frac{x - E_{\rm min}}{E_{\rm max} - E_{\rm min}}
\end{equation}
where $E_{\rm min} = {\it min}({\bf E}), E_{\rm max} = {\it max}({\bf E}), {\bf E} = H({\bf Q})$.

\subsection{Scaling Functions}

The scaling function {\ScaleF} serves two purposes: (i) to \emph{invert} the normalised entropy value, such that low entropy positions are given a high value in the weighting vector, and (ii) to \emph{scale} the differential between a high weighting and a low weighting.

We considered four types of scaling functions:
\begin{itemize}
\item A hyperbolic scaler of the form:
\begin{equation}
\ScaleF_{\rm hyper}(x) = \frac{1}{(1 + ax)^c}
\end{equation}
where $a > 0$ and $c > 0$ are scaling constants.

\item An exponential scaler of the form:
\begin{equation}
\ScaleF_{\rm exp}(x) = e^{-kx}
\end{equation}
where $k > 0$ is a scaling constant.

\item A sigmoid scaler of the form:
\begin{equation}
\ScaleF_{\rm sigmoid}(x) = \frac{1}{1 + e^{k(x - \tau)}}
\end{equation}
where $k > 0$ is a scaling constant $0 \leq \tau \leq 1$ is the threshold constant.

\item A thresholding step function:
\begin{equation}
\ScaleF_{\rm thresh}(x) = 
\begin{cases}
1 & \text{if}\ x \leq \tau \\
0 & \text{otherwise} 
\end{cases}
\end{equation}
where $0 \leq \tau \leq 1$ is the thresholding constant.
\end{itemize}

\subsection{Entropy Weighted Needleman-Wunsch}

We propose a modified Needleman-Wunsch scoring matrix $F^{*}$
(replacing $F$). The initialisation
and traceback steps for $F^{*}$ follow the same method
as for standard Needleman-Wunsch (see \cf~\ref{ss:nw}.) However for the progressive scoring
calculation, weights are applied from the vector ${\bf w}$:

\vspace{-0.4cm}
\begin{equation}
F^{*}_{i,j} = {\it max}
	\left\{ \begin{array}{l}
		F^{*}_{i-1,j-1} + w_k\! \cdot\! S(M_1(i), M_2(j)) \\
		F^{*}_{i,j-1} + w_k\! \cdot\! d_{\rm gap} \\
		F^{*}_{i-1,j} + w_k\! \cdot\! d_{\rm gap} \\
	\end{array}
	\right.
\end{equation}
where $k = {\it max}\{i, j\}$.

%\begin{equation}
%S^{*}(M_{1_i},M_{2_j}) = 
%\begin{cases}
%w_i\! \cdot\! d_{\rm identical}  &~\text{if}~M_{1_i} = M_{2_j} \\
%w_i\! \cdot\! d_{\rm differing} &~\text{otherwise}\\
%\end{cases}
%\end{equation}

\begin{figure}
\includegraphics[width=.4\textwidth]{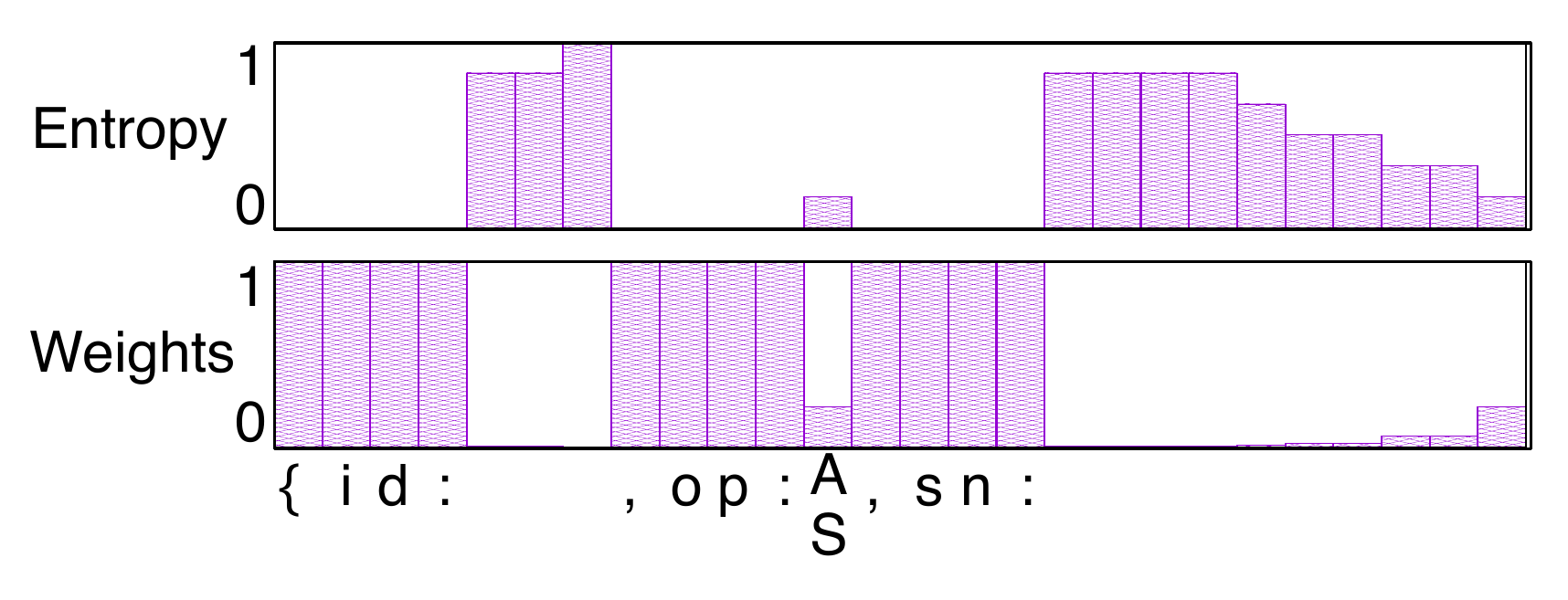}
\caption{Example entropy and weights calculated from the interaction library in Table~\ref{tab:tl},
using the Richness and hyperbolic scaler ($a=1, c=10$)
}
\label{fig:weightsexample}
\end{figure}

Figure~\ref{fig:weightsexample} shows an example entropy and weights vectors
calculated from the interaction library in Table~\ref{tab:tl} (using the
Richness entropy method, and the hyperbolic scaler with $a=1, c=10$). Returning to
our example from Section~\ref{ss:noweightexample}, by applying the entropy
weighting method, a different response is selected. Using the weights vector
from Figure~\ref{fig:weightsexample}, request 4 (\DistFunc=0.0066) now has a
closer distance to the live request than request 3 (\DistFunc=0.019), thereby
selecting a response of the correct operation type.

\section{Experimental Evaluation}
\label{sec:eval}

We wanted to evaluate the entropy-weighted response selection to answer the
following research questions:

\begin{enumerate}
\item \textbf{RQ1} (\textbf{Comparison of Entropy Methods}):
what impact do the differing entropy methods have on the response
accuracy?
\item \textbf{RQ2} (\textbf{Comparison of Scaling Functions}):
what is the impact of the scaling function on the response
accuracy and the sensitivity of their parameters?
\item \textbf{RQ3} (\textbf{Relative Accuracy}):
what is the overall accuracy of the entropy weighted response
selection relative to other methods?
\end{enumerate}

\subsection{Case Study Protocols and Traces}
\label{ss:casestudies}

%%% NOTE: we may have to check these numbers against the actual traces... %%%
\begin{table}[h]
\centering
\resizebox{0.49\textwidth}{!}{
\begin{tabular}{|l|c|c|c|c|}
  \hline
  Protocol & Binary/Text & Fields & \#Ops. & \#Transactions \\
  \hline \hline
  IMS & binary & fixed length & 5 & 800 \\ \hline
  LDAP & binary & length-encoded & 10 & 2177 \\ \hline
  SOAP & text & delimited & 6 & 1000 \\ \hline
  Twitter (REST) & text & delimited & 6 & 1825 \\ \hline
\end{tabular}}
\caption{Message trace datasets. % (available at~\cite{datasets})
}
\label{tab:sampleset}
\end{table}

In order to answer these questions, we applied our technique on message
trace datasets from four case study protocols:
\emph{IMS}~\cite{imsredbook} (a binary mainframe protocol),
\emph{LDAP}~\cite{ldap} (a binary directory service protocol),
\emph{SOAP}~\cite{soap} (a textual protocol, with an Enterprise Resource
Planning (ERP) system messaging system services), and
\emph{RESTful Twitter}~\cite{twitter} (a JSON protocol for the Twitter social
media service). We chose these four protocols because: (i) they are widely used in enterprise environments, (ii) they
represent a good mix of text-based protocols (SOAP and RESTful Twitter) and
binary protocols (IMS and LDAP), (iii) they use either fixed length, length encoding or
delimiters to structure protocol messages, %\footnote{Given a protocol message,
%  length fields or delimiters are used to convert its structure into a
%  sequence of bytes that can be transmitted over the wire. Specifically, a
%  length field is a number of bytes that show the length of another field,
%  while a delimiter is a byte (or byte sequence) with a
%  known value that indicates the end of a field.}
and (iv) each of them includes a diverse
number of operation types, as indicated by the \emph{Ops} column. The number
of request-response interactions for each test case is shown as column
\emph{\#Transactions} in Table~\ref{tab:sampleset}.

Our message trace datasets are available for download
% from our website~\cite{datasets}.
at {\sf {\small { http://quoll.ict.swin.edu.au/doc/message\_traces.html}}}

\begin{table}[t]
\begin{center}
\resizebox{0.49\textwidth}{!}{
\begin{tabular}{|l|l|}
\hline
Incoming Request & \{id:15,op:S,sn:Du\} \\
\hline
Expected Response & \{id:15,op:SearchRsp,result:Ok,gn:Miao,sn:Du\} \\
\hline
\multirow{2}{*}{Valid Responses} & \{id:15,op:SearchRsp,result:Ok,gn:Miao,sn:Du\} \\\cline{2-2}
& \{id:15,op:SearchRsp,result:Ok,gn:Menka,sn:Du\} \\
\hline
\multirow{2}{*}{Invalid Responses} & \{id:15,op:\textbf{\emph{AddRsp}},result:Ok\} \\\cline{2-2}
& \{id:15,op:SearchRsp,result:Ok,gn:Miao\textbf{\emph{\},sn:Du}} \\
\hline
\end{tabular}
}
\end{center}
\caption{Examples of valid and invalid emulated responses.}
\label{tab:validexample}
\end{table}

\subsection{Compared Techniques}

We compared the proposed entropy-weighted response selection with two other methods. The
baseline for comparison was a hash lookup. If the hash code of an incoming
request matched the hash code of a request in the transaction library, then
the associated response was replayed (without any transformation). This
approach can only work when a request is identical to the live request occurred
in the recording. It is a standard record-and-replay approach used for
situations where nothing is known about the protocol. Our second compared
technique is the non-weighted Needleman-Wunsch response selection~\cite{Du:2013}.

\subsection{Accuracy Measurement}

%%% Removed after discussions this morning... %%%
% \begin{figure}[h]
% \centering
% \includegraphics[width=7cm, height=4.25cm]{evaluationprocess}
% \caption{10-fold Cross Validation Approach}
% \label{fig:evaluationprocedure}
% \end{figure}

Cross-validation %\cite{devijver:1982}
is a popular model validation
method for assessing how accurately a predictive model will perform in
practice. For the purpose of our evaluation, we applied the commonly used
10-fold cross-validation approach \cite{mclachlan:2004} to
all four case study datasets.

% As shown in Figure \ref{fig:evaluationprocedure},
We randomly partitioned each of the original interaction datasets into 10 groups. Of
these 10 groups, one group
% (the top-left rectangle in Figure~\ref{fig:evaluationprocedure})
is considered to be the {\em evaluation group} for testing our approach, and
the remaining 9 groups constitute the {\em training set}. This process is then
repeated 10 times (the same as the number of groups), so that each of the 10
groups will be used as the evaluation group once.
When running each experiment with each trace dataset, we applied our approach to each request message in
the {\em evaluation group}, referred to as the {\em incoming request}, to generate an
{\em emulated response}.
The entire cross validation process was repeated ten times for each
experiment using different random seeds.
%We recorded the time that our approach took to generate the response for each
%incoming request. This was used to evaluate the runtime efficiency of our
%approach.
%And then, we compared the resulting {\em generated response} with the recorded({\em expected response}) of the {\em incoming request}. For the result of the comparison,

Having generated a response for each incoming request, we then
used a validation script to assess the accuracy.
The script used a protocol decoder
to parse the emulated response and compared it to the original
recorded response (the \emph{expected response}) from the evaluation group.
The emulated response was classified as \emph{valid} or \emph{invalid}, according to the following
definitions:
\begin{enumerate}
\item \textbf{\em Valid:} the emulated response conformed to the message format
of the protocol (\ie was successfully parsed by the decoder)
and the operation type of the emulated response was the same as the expected response.
Note that contents of the emulated response payload may differ to the
expected response and still be considered valid.
\item \textbf{\em Invalid:} the emulated response was not structured according
to the expected message format of the protocol, or the operation type of
the emulated response was different to the expected response.
\end{enumerate}

\subsection{Entropy Method Results (RQ1)}

The three alternative entropy measure functions were used against the four
datasets. Keeping the scaling method consistent (Hyperbolic scaler, $a=50, c=1$),
the accuracy of the different methods is shown in Table~\ref{tab:emethodres}.
The Richness method was the best for the IMS dataset, and the Shannon Index
was the best for LDAP. All of the entropy methods outperformed the
non-weighted approach.

\begin{table}
\begin{center}
\begin{tabular}{|c|c|c|c|c|}
\hline
Entropy & \multirow{2}{*}{IMS} & \multirow{2}{*}{LDAP} & \multirow{2}{*}{SOAP} & \multirow{2}{*}{Twitter} \\
Method & & & & \\
\hline
None & 77.4\%             & 94.2\%          & 100\% & 99.5\% \\
Shannon & 81.3\%          & \textbf{94.9\%} & 100\% & 99.5\% \\
Richness & \textbf{100\%} & 91.6\%          & 100\% & 99.5\% \\
Simpson & 89.9\%          & 94.3\%          & 100\% & 99.5\% \\
\hline
\end{tabular}
\caption{The response accuracy for the alternative entropy measures against the four datasets tested.}
\label{tab:emethodres}
\end{center}
\end{table}

\subsection{Scaling Function Results (RQ2)}

\begin{figure*}[ht]
\centering
\begin{minipage}[b]{0.33\textwidth}
\includegraphics[width=.9\textwidth]{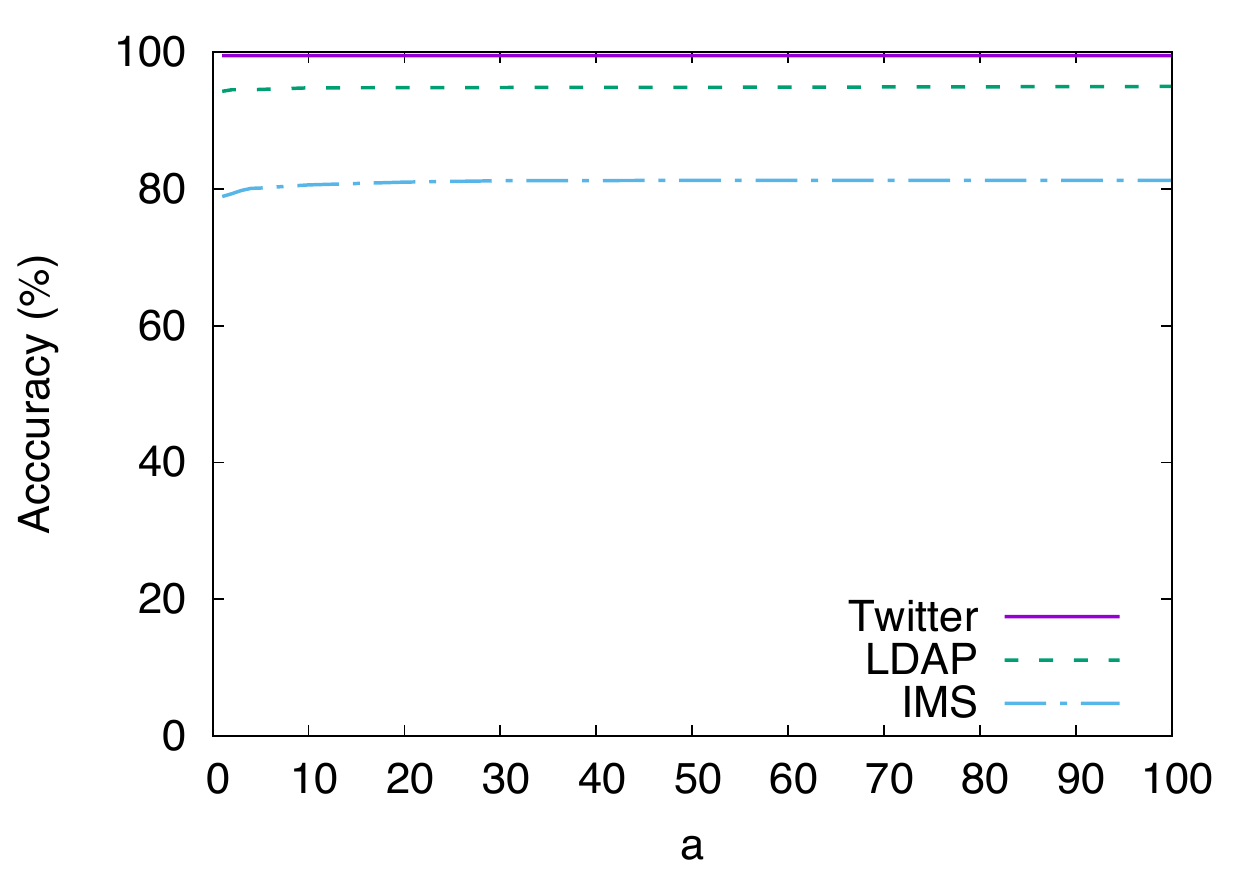}\\
\vspace{-.5cm} 
\subcaption{\small Hypberbolic scaler: $a$ sensitivity}
\end{minipage}%
\begin{minipage}[b]{0.33\textwidth}
\includegraphics[width=.9\textwidth]{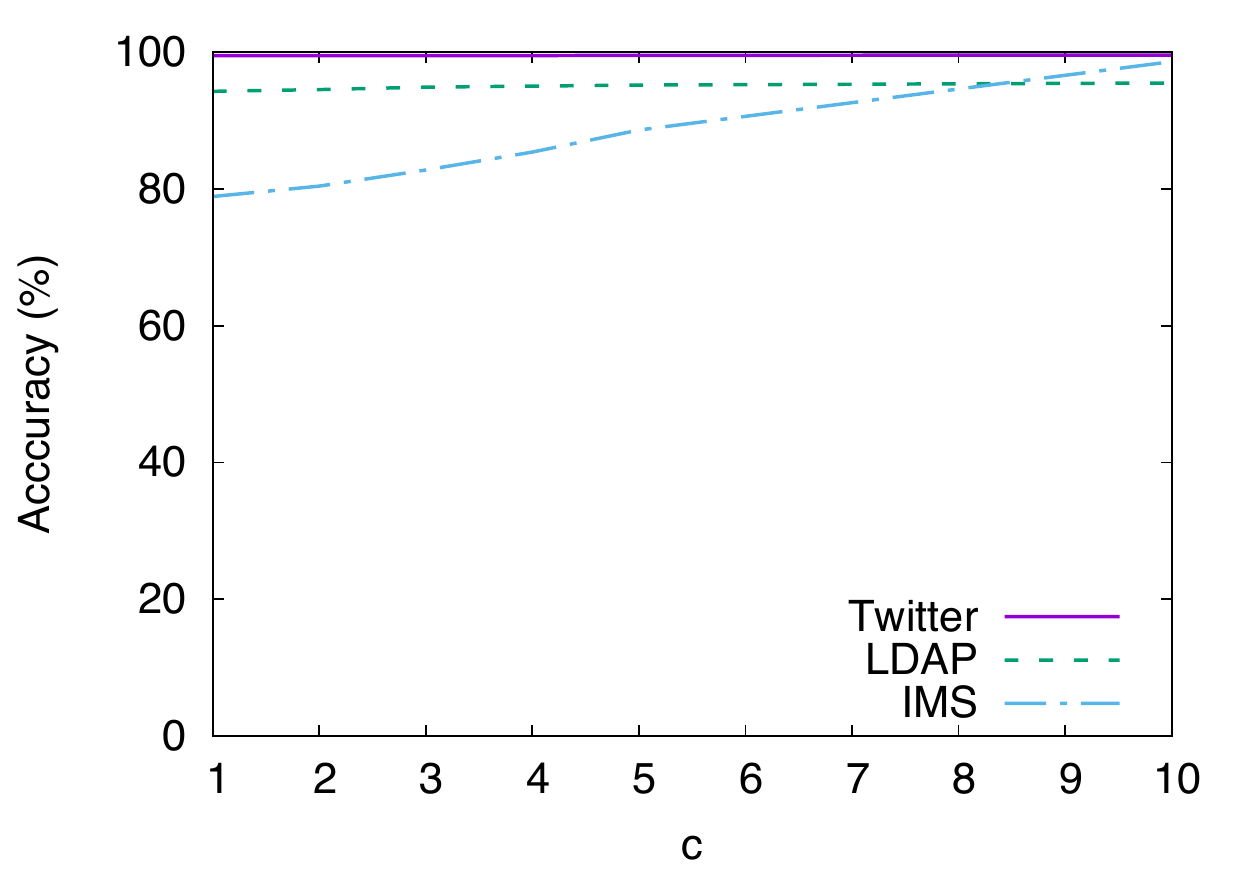}\\
\vspace{-.5cm} 
\subcaption{\small Hyperbolic scaler: $c$ sensitivity}
\end{minipage}%
\begin{minipage}[b]{0.33\textwidth}
\includegraphics[width=.9\textwidth]{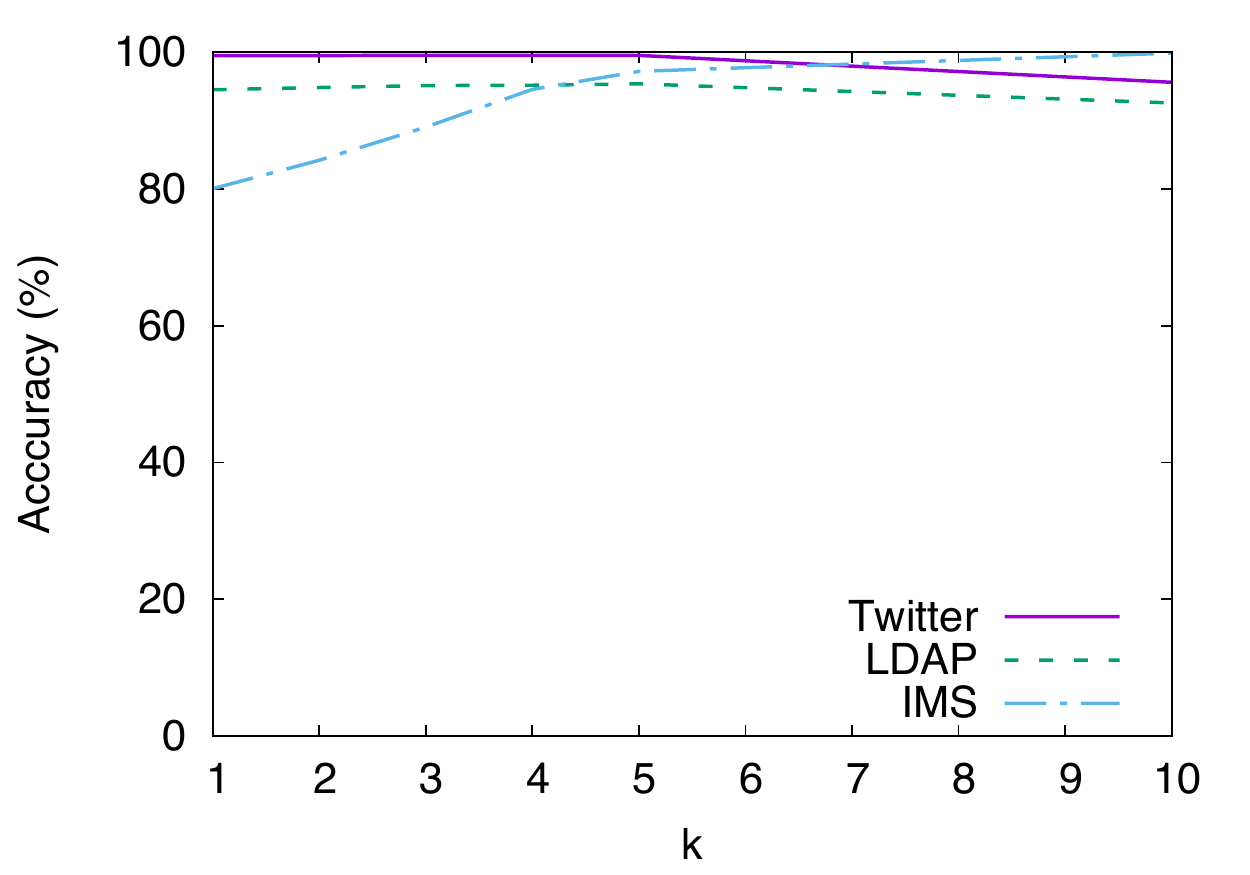}\\
\vspace{-.5cm} 
\subcaption{\small Exponential scaler: $k$ sensitivity}
\end{minipage}%
\\
\begin{minipage}[b]{0.3333\textwidth}
\includegraphics[width=.9\textwidth]{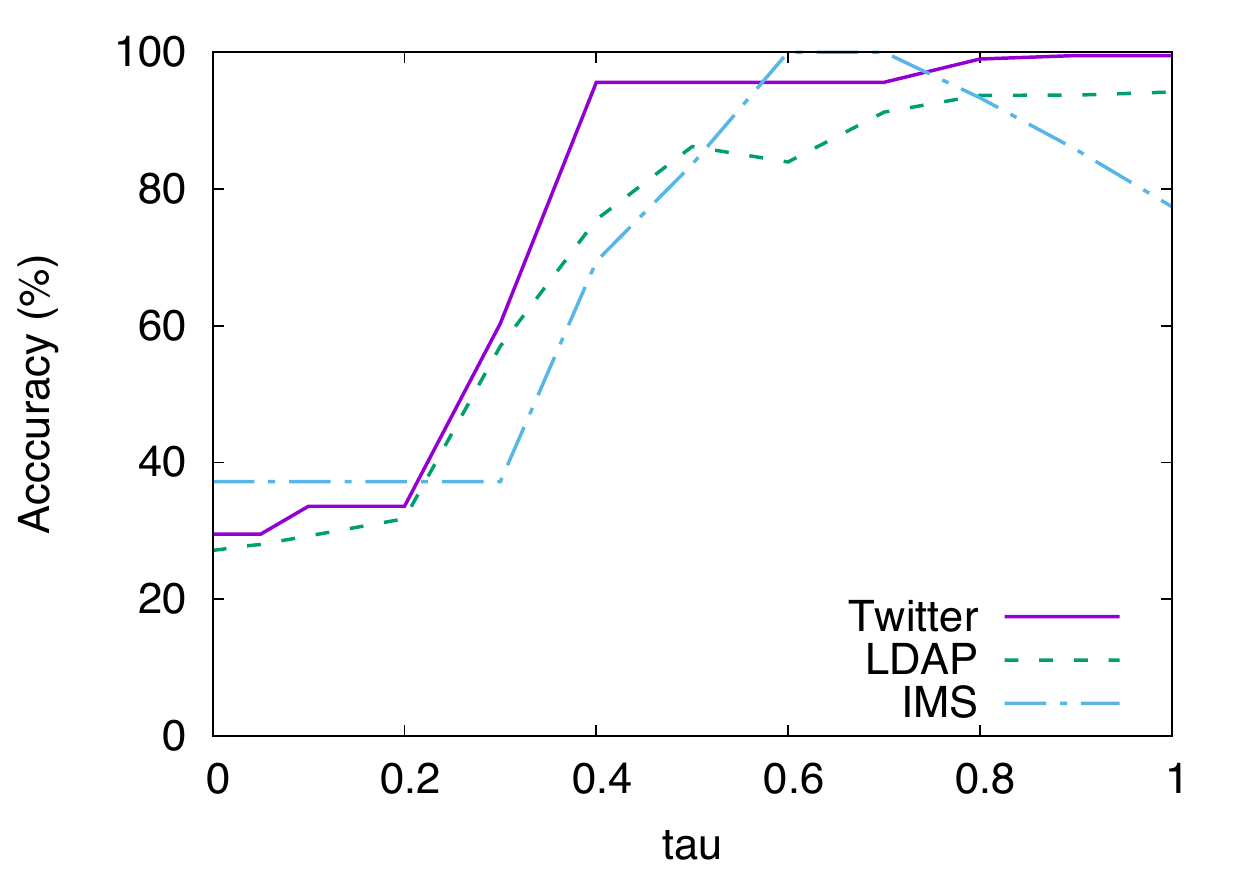}\\
\vspace{-.5cm} 
\subcaption{\small Thresholding scaler: $\tau$ sensitivity}
\end{minipage}%
\begin{minipage}[b]{0.33\textwidth}
\includegraphics[width=.9\textwidth]{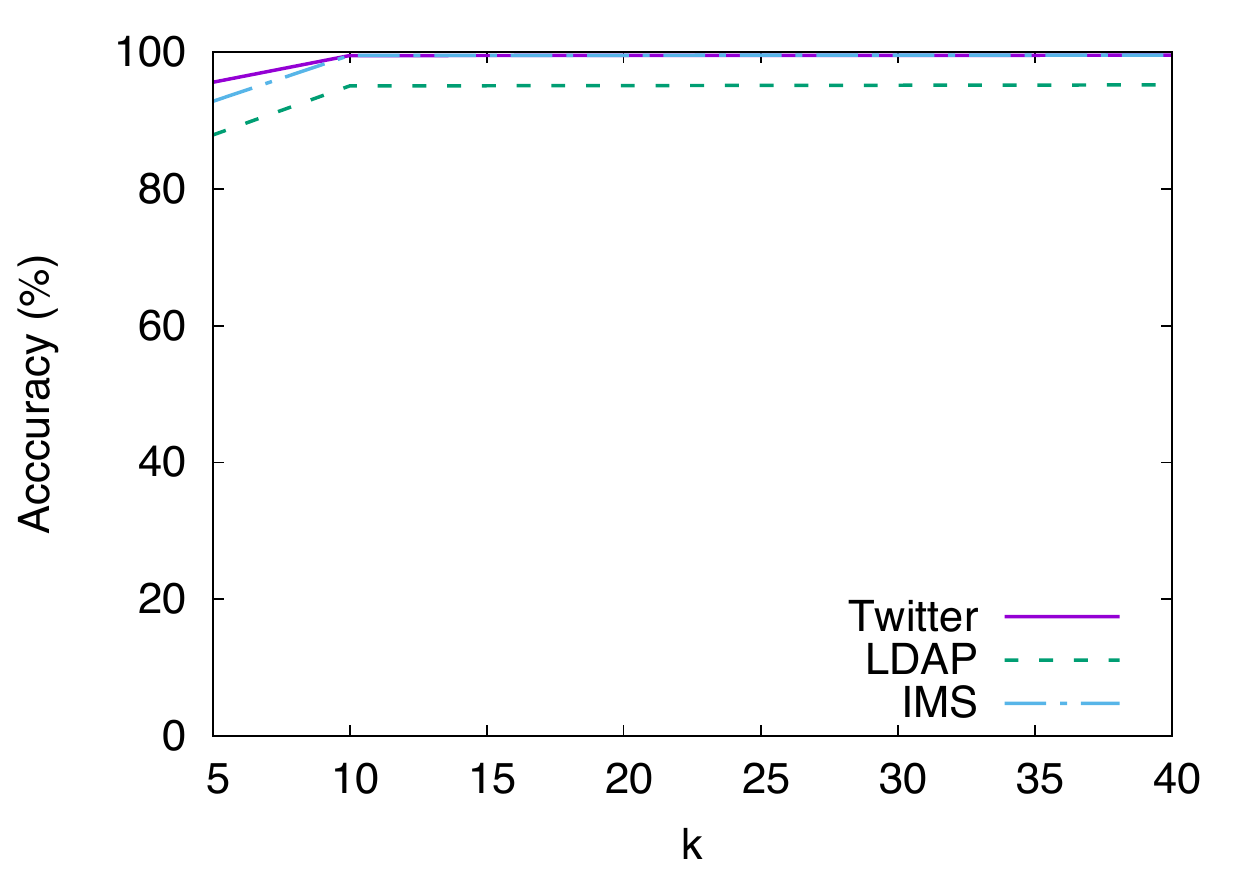}\\
\vspace{-.5cm} 
\subcaption{\small Sigmoid scaler: $k$ sensitivity}
\end{minipage}%
\begin{minipage}[b]{0.33\textwidth}
\includegraphics[width=.9\textwidth]{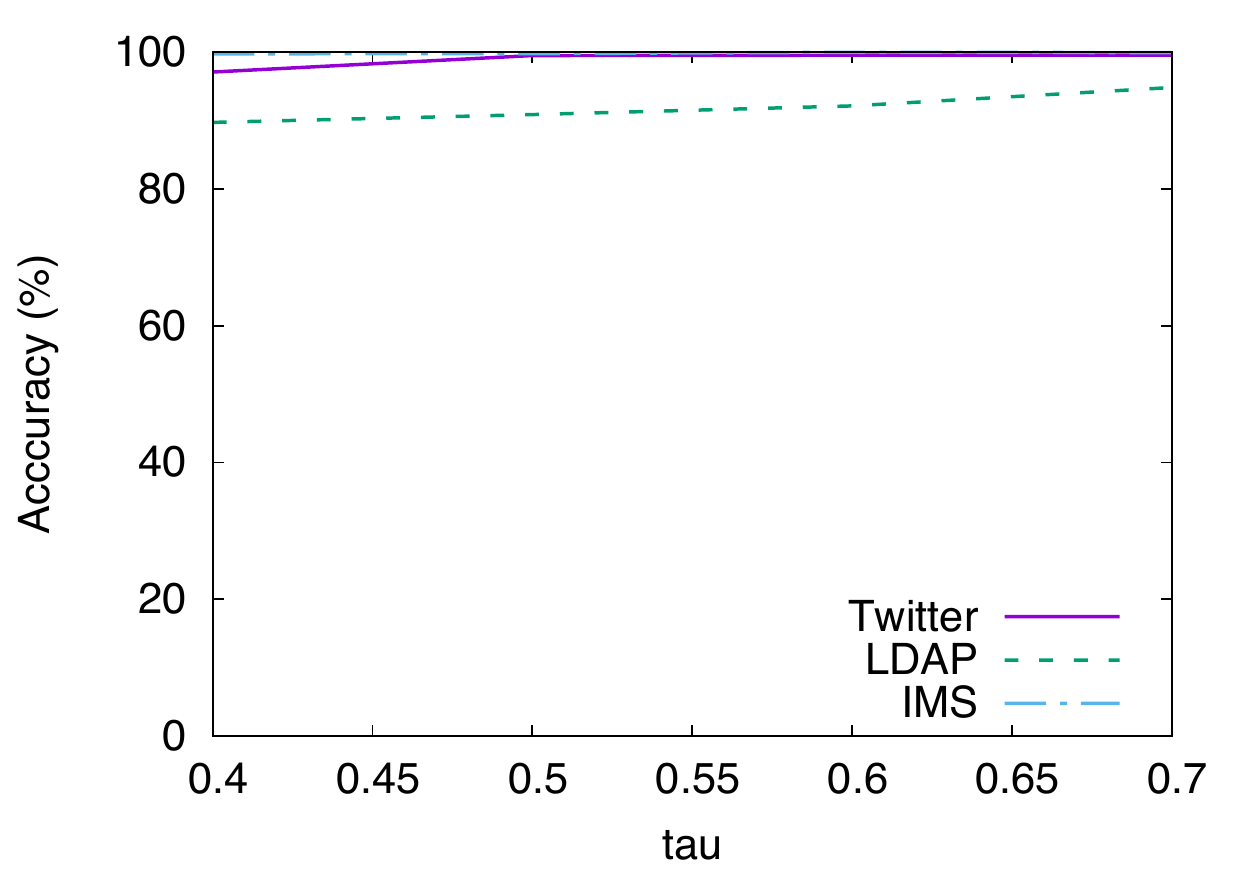}\\
\vspace{-.5cm} 
\subcaption{\small Sigmoid scaler: $\tau$ sensitivity}
\end{minipage}%\
\caption{Effect of scaling function parameters on response accuracy}
\label{fig:scalingres}
\end{figure*}

The scaling functions were applied with varying parameters, keeping the
entropy method consistent (Shannon Index). The results are shown in
Figure~\ref{fig:scalingres}. For the Hyperbolic scaler, the most significant
results can be observed for IMS, where the accuracy increases from 80\% to
99\% as the $c$ scaling exponent is increased. Similarly for the Exponential
scaler, accuracy for IMS increases with a higher exponent value. For the other
datasets no dramatic differences are observed. The Thresholder scaler is very
sensitive to the setting of the threshold $\tau$, and the optimal setting is
different for each data set. The performance of the Sigmoid scaler appears
fairly insensitive to the parameter settings.

%%% One could fiddle around a bit more with gaps here... (JGS) %%%

\subsection{Comparison to Baselines (RQ3)}

Our results showed that the Shannon Index combined with the Hypberbolic scaler
($a=1, c=10$) gave high response accuracy for all datasets.
Table~\ref{tab:baseline} compares this method to the other response selection
approaches.  The entropy weighted method equalled or outperformed the other
methods for all datasets. The most significant improvement occurs with IMS.
There is a marginal improvement for LDAP and Twitter. SOAP already had 100\%
accuracy using the non-weighted method, but importantly the entropy weighting
did not worsen this result.

\begin{table}[t]
\centering
\begin{tabular}{|l|c|c|c|c|}
\hline
Matching Method & IMS & LDAP & SOAP & Twitter \\
\hline
Hash Lookup               & 50\%            & 5.36\%          & 0.5\% & 30.1\% \\
Non-Weighted NW & 77.4\%          & 94.2\%          & \textbf{100\%} & 99.5\% \\
Entropy-Weight NW & \textbf{98.6\%} & \textbf{95.5\%} & \textbf{100\%} & \textbf{99.6\%} \\
\hline
\end{tabular}
\caption{The response accuracy of entropy weighted matching versus other response selection methods}
\label{tab:baseline}
\end{table}

\section{Discussion and Future Work}

In the experiments, the entropy weighted method had a significant impact on
improving the response accuracy for the IMS dataset, but had limited impact
for the other three datasets. A key characteristic of IMS is that it uses
fixed length encoding to delimit fields. This ensures the operation type
will always be at the same character position. For the other datasets, the
operation type can be at different positions in the message. We hypothesise that
this is the cause of the decreased effectiveness for non-fixed width protocols.
Further testing is required to confirm this.

Regarding the entropy measures tested, no one method was conclusively
better. However we caution that Richness will be more sensitive to
noisy data (with single data points distorting the results) whereas
the Shannon and Simpson indices are more robust.

Future work will consider aligning all the messages in the interaction library
before performing entropy analysis (using multiple sequence alignment). We
expect this to improve results for variable width protocols, by increasing
the probability that the operation type characters are aligned.
Another extension is to cluster the interaction library first, and
then perform a separate entropy analysis on each cluster.

\section{Conclusion}

Service virtualisation is an important tool for realising continuous delivery,
by creating realistic service models of a system-under-test's dependency
services, thereby facilitating automated testing of production-like
conditions. Opaque service virtualisation is a method for automatically
deriving service models, even in the absence of a protocol decoder and other
knowledge.
We perform an entropy analysis on a recorded sample of the target service's messages
and then use an entropy weighted Needleman-Wunsch based similarity measure to select the
best matching responses to play back to live requests.
We have shown our entropy weighted approach can be used to improve the accuracy
of opaque service virtualisation, particularly for fixed width protocols.

%ACKNOWLEDGMENTS are optional
\section{Acknowledgments}
This research was supported by ARC grant LP150100892.
%This section is optional; it is a location for you
%to acknowledge grants, funding, editing assistance and
%what have you.  In the present case, for example, the
%authors would like to thank Gerald Murray of ACM for
%his help in codifying this \textit{Author's Guide}
%and the \textbf{.cls} and \textbf{.tex} files that it describes.

\bibliographystyle{abbrv}
\bibliography{./VersteegEntropy2016}  % sigproc.bib is the name of the Bibliography in this case
\end{document}